\documentstyle[aps,twocolumn]{revtex}
\begin{document}
\draft
\title{Giant backscattering peak in angle-resolved Andreev reflection}
\author{C. W. J. Beenakker, J. A. Melsen, and P. W. Brouwer}
\address{Instituut-Lorentz, University of Leiden,
P.O. Box 9506, 2300 RA Leiden, The Netherlands\medskip\\
\parbox{14cm}{\rm
It is shown analytically and by numerical simulation that the angular
distribution of Andreev reflection by a disordered
normal-metal--superconductor junction has a narrow peak at the angle of
incidence. The peak is higher than the well-known coherent
backscattering peak in the normal state, by a large factor $G/G_{0}$
(where $G$ is the conductance of the junction and $G_{0}=2e^{2}/h$).
The enhanced backscattering can be detected by means of ballistic point
contacts.\smallskip
--- {\sf cond-mat/9501003}}}
\maketitle
\narrowtext

Coherent backscattering is a fundamental effect of time-reversal symmetry on
the reflection of electrons by a disordered metal \cite{Alt92,Ber94}. The
angular reflection distribution has a narrow peak at the angle of incidence,
due to the constructive interference of time-reversed sequences of multiple
scattering events. At zero temperature, the peak is twice as high as the
background. Coherent backscattering manifests itself in a transport experiment
as a small negative correction of order $G_{0}=2e^2/h$ to the average
conductance $G$ of the metal ({\em weak localization\/} \cite{Ber84}). Here we
report the theoretical prediction, supported by numerical simulations, of a
giant enhancement of the backscattering peak if the normal metal (N) is in
contact with a superconductor (S). At the NS interface an electron incident
from N is reflected either as an electron (normal reflection) or as a hole
(Andreev reflection). Both scattering processes contribute to the
backscattering peak. Normal reflection contributes a factor of two. In
contrast, we find that Andreev reflection contributes a factor $G/G_{0}$, which
is $\gg 1$.

If the backscattering peak in an NS junction is so large, why has it not been
noticed before in a transport experiment? The reason is a cancellation in the
integrated angular reflection distribution which effectively eliminates the
contribution from enhanced backscattering to the conductance of the NS
junction. However, this cancellation does not occur if one uses a ballistic
point contact to inject the current into the junction. We discuss two
configurations, both of which show an excess conductance due to enhanced
backscattering which is a factor $G/G_{0}$ greater than the weak-localization
correction.

We consider a disordered normal-metal conductor (length $L$, width $W$, mean
free path $l$, with $N$ propagating transverse modes at the Fermi energy
$E_{\rm F}$) which is connected at one end to a superconductor (see inset of
Fig.\ 1). An electron (energy $E_{\rm F}$) incident from the opposite end in
mode $m$ is reflected into some other mode $n$, either as an electron or as a
hole, with probability amplitudes $r_{nm}^{\rm ee}$ and $r_{nm}^{\rm he}$,
respectively. The $N\times N$ matrices $r^{\rm ee}$ and $r^{\rm he}$ are given
by \cite{Bee92}
\begin{mathletters}
\label{rsrelation}
\begin{eqnarray}
r^{\rm ee}&=&s_{11}^{\vphantom{\ast}}- s_{12}^{\vphantom{\ast}}s_{22}^{\ast}
(1+s_{22}^{\vphantom{\ast}}s_{22}^{\ast})^{-1}
s_{21}^{\vphantom{\ast}}, \label{ree}\\
r^{\rm he}&=&-{\rm i}s_{12}^{\ast}
(1+s_{22}^{\vphantom{\ast}}s_{22}^{\ast})^{-1}
s_{21}^{\vphantom{\ast}}. \label{rhe}
\end{eqnarray}
\end{mathletters}%
The $s_{ij}$'s are submatrices of the scattering matrix $S$ of the disordered
normal region,
\[
S={\renewcommand{\arraystretch}{0.6}
\left(\begin{array}{cc}
s_{11}&s_{12}\\s_{21}&s_{22}
\end{array}\right)}=
{\renewcommand{\arraystretch}{0.6}
\left(\begin{array}{cc}
u&0\\0&v
\end{array}\right)}
{\renewcommand{\arraystretch}{0.6}
\left(\begin{array}{cc}
-\sqrt{\cal R}&\sqrt{\cal T}\\
\sqrt{\cal T}&\sqrt{\cal R}
\end{array}\right)}
{\renewcommand{\arraystretch}{0.6}
\left(\begin{array}{cc}
u'&0\\0&v'
\end{array}\right)},
\]
where $u,v,u',v'$ are $N\times N$ unitary matrices, ${\cal R}=1-{\cal T}$, and
$\cal T$ is a diagonal matrix with the transmission eigenvalues
$T_{1},T_{2},\ldots T_{N}$ on the diagonal.

We first consider zero magnetic field ($B=0$). Time-reversal symmetry then
requires that $S$ is a symmetric matrix, hence $u'=u^{\rm T}$, $v'=v^{\rm T}$.
Eq.\ (\ref{rsrelation}) simplifies to
\begin{equation}
r^{\rm ee}=-2u\frac{\sqrt{1-{\cal T}}}{2-{\cal T}}u^{\rm T},\;\;
r^{\rm he}=-{\rm i}u^{\ast}\frac{\cal T}{2-{\cal T}}u^{\rm T}.
\label{rUrelation}
\end{equation}
We seek the average reflection probabilities $\langle|r_{nm}|^{2}\rangle$,
where $\langle\cdots\rangle$ denotes an average over impurity configurations.
Following Mello, Akkermans, and Shapiro \cite{Mel88}, we assume that $u$ is
uniformly distributed over the unitary group. This ``isotropy assumption'' is
an approximation which ignores the finite time scale of transverse diffusion.
The reflection probabilities contain a product of four $u$'s, which can be
averaged by means of the formula \cite{Mel90}
\begin{eqnarray}
\langle u^{\vphantom{\ast}}_{ni}u^{\vphantom{\ast}}_{mj}
u^{\ast}_{nk}u^{\ast}_{ml}\rangle&=&
(N^{2}-1)^{-1}(\delta_{ik}\delta_{jl}+
\delta_{nm}\delta_{il}\delta_{jk})-\nonumber\\
&&\!\!(N^{3}-N)^{-1}(\delta_{il}\delta_{jk}+
\delta_{nm}\delta_{ik}\delta_{jl}).
\label{U4average}
\end{eqnarray}
The result is (with the definition $\tau_{k}\equiv T_{k}(2-T_{k})^{-1}$)
\begin{mathletters}
\label{rnmT}
\begin{eqnarray}
\langle|r_{nm}^{\rm ee}|^{2}\rangle&=&\frac{\delta_{nm}+1}{N^{2}+N}\left(N-
\langle{\textstyle\sum_{k}}\tau_{k}^{2}\rangle\right),\label{rnmTee}\\
\langle|r_{nm}^{\rm he}|^{2}\rangle&=&
\frac{\delta_{nm}+1}{N^{2}+N}\langle{\textstyle\sum_{k}}\tau_{k}^{2}\rangle
+\frac{N\delta_{nm}-1}{N^{3}-N}\langle{\textstyle\sum_{k\neq
k'}}\tau_{k}\tau_{k'}\rangle.\nonumber\\
\label{rnmThe}
\end{eqnarray}
\end{mathletters}%
In the metallic regime $N\gg L/l\gg 1$. In this large-$N$ limit we may
factorize $\langle\sum_{k\neq k'}\tau_{k}\tau_{k'}\rangle$ into
$\langle\sum_{k}\tau_{k}\rangle^{2}$, which can be evaluated using \cite{Sto91}
\begin{equation}
\langle{\textstyle\sum_{k}}f(T_{k})\rangle=(Nl/L){\textstyle\int_{0}^{\infty}}
dx\,f(1/\cosh^{2}x).\label{fTaverage}
\end{equation}
The result for normal reflection is
\begin{equation}
\langle|r_{nm}^{\rm
ee}|^{2}\rangle=(1+\delta_{nm})N^{-1}(1-\case{1}{2}l/L).\label{normal}
\end{equation}
Off-diagonal ($n\neq m$) and diagonal ($n=m$) reflection differ by precisely a
factor of two, just as in the normal state \cite{Mel88}.
In contrast, for Andreev reflection we find
\begin{equation}
\langle|r_{nm}^{\rm he}|^{2}\rangle=\case{1}{2}l/NL \;\;(n\neq m),\;\;
\langle|r_{nn}^{\rm he}|^{2}\rangle=(\pi l/4L)^{2}.\label{Andreev}
\end{equation}
Off-diagonal and diagonal reflection now differ by an order of magnitude
$Nl/L\simeq G/G_{0}\gg 1$.

Eqs.\ (\ref{normal}) and (\ref{Andreev}) hold for $B=0$. If time-reversal
symmetry is broken (by a magnetic field $B\gtrsim B_{\rm c}\equiv h/eLW$), then
the matrices $u,u',v,v'$ are all independent \cite{Sto91}. Carrying out the
average in the large-$N$ limit, we find
\begin{equation}
\langle|r_{nm}^{\rm ee}|^{2}\rangle=N^{-1}(1-\case{1}{2}l/L),\;\;
\langle|r_{nm}^{\rm he}|^{2}\rangle=\case{1}{2}l/NL.\label{noTRS}
\end{equation}
Diagonal and off-diagonal reflection now occur with the same probability.

We have checked this theoretical prediction of a giant backscattering peak by a
numerical simulation along the lines of Ref.\ \cite{Mar93}. The disordered
normal region was modeled by a tight-binding Hamiltonian on a two-dimensional
square lattice (dimensions $300\times 300$, $N=126$), with a random impurity
potential at each site ($L/l=9.5$). The scattering matrix $S$ was computed
numerically and then substituted into Eq.\ (\ref{rsrelation}) to yield $r^{\rm
ee}$ and $r^{\rm he}$. Results are shown in Fig.\ 1. This is raw data from a
single sample. For normal reflection (bottom panel) the backscattering peak is
not visible due to statistical fluctuations in the reflection probabilities
(speckle noise). The backscattering peak for Andreev reflection is much larger
than the fluctuations and is clearly visible (top panel). A magnetic flux of
$10\,h/e$ through the disordered region completely destroys the peak (middle
panel). The arrow in the top panel indicates the ensemble-averaged peak height
from Eq.\ (\ref{Andreev}), consistent with the simulation within the
statistical fluctuations. The peak is just one mode wide, as predicted by Eq.\
(\ref{Andreev}). If $W>L$ the isotropy assumption breaks down \cite{Mel88} and
we expect the peak to broaden over $W/L$ modes. Fig.\ 1 tells us that for $L=W$
the isotropy assumption is still reasonably accurate in this problem.

Coherent backscattering in the normal state is intimately related to the
weak-localization correction to the average conductance. We have found that the
backscattering peak for Andreev reflection is increased by a factor $G/G_{0}$.
However, the weak-localization correction in an NS junction remains of order
$G_{0}$ \cite{Bee92,Tak94}. The reason is that the conductance
\begin{equation}
G=2G_{0}{\textstyle\sum_{n,m}}|r_{nm}^{\rm he}|^{2}\label{GNS}
\end{equation}
contains the sum over all Andreev reflection probabilities \cite{Tak92}, so
that the backscattering peak is averaged out. Indeed, Eqs.\ (\ref{Andreev}) and
(\ref{noTRS}) give the same $G$, up to corrections smaller by factors $1/N$ and
$l/L$. In order to observe the enhanced backscattering in a transport
experiment one has to increase the sensitivity to Andreev reflection at the
angle of incidence. This can be done by injecting the electrons through a
ballistic \cite{Vol94} point contact (width $\ll l$, number of transmitted
modes $N_{0}$). For $B=0$, one can compute the average conductance from
\cite{Bee92}
\begin{equation}
\langle G\rangle=2G_{0}{\textstyle\int_{0}^{1}}dT\,\rho(T)T^{2}(2-T)^{-2}.
\label{GNSnul}
\end{equation}
The density of transmission eigenvalues $\rho(T)$ is known \cite{Bee94,Naz94},
in the regime $N_{0}\gg 1$, $N\gg L/l$. One finds
\begin{mathletters}
\label{GBzero}
\begin{eqnarray}
&&\langle G\rangle=G_{0}[\case{1}{2}(1+\sin\vartheta)/N_{0}+L/Nl\,]^{-1},
\label{GBzeroa}\\
&&\case{1}{2}\vartheta(1+\sin\vartheta)=(N_{0}L/Nl)\cos\vartheta,\;\;
\vartheta\in(0,\pi/2).\label{GBzerob}
\end{eqnarray}
\end{mathletters}%
In the absence of time-reversal symmetry ($B\gtrsim B_{\rm c}$) we find from
the large-$N$ limit of Eqs.\ (\ref{rsrelation}) and (\ref{GNS}) that
\begin{equation}
\langle G\rangle=G_{0}(1/N_{0}+L/Nl)^{-1}.\label{GBc}
\end{equation}
This is just the classical addition in series of the Sharvin conductance
$N_{0}G_{0}$ of the point contact and the Drude conductance $(Nl/L)G_{0}$ of
the disordered region.

In Fig.\ 2 we have plotted the difference $\Delta G=\langle
G(B=0)\rangle-\langle G(B\gtrsim B_{\rm c})\rangle$ of Eqs.\ (\ref{GBzero}) and
(\ref{GBc}). If $N_{0}/N\ll l/L\ll 1$ the conductance drops from $2N_{0}G_{0}$
to $N_{0}G_{0}$ upon breaking time-reversal symmetry. A doubling of the contact
conductance at $B=0$ is well-known \cite{Son87} in {\em ballistic\/} NS
junctions ($l\gg L$). There it has a simple classical origin: An electron
injected towards the NS interface is reflected back as a hole, doubling the
current through the point contact. Here we find that the conductance doubling
can survive multiple scattering by a disordered region ($l\ll L$), as a result
of enhanced backscattering at the angle of incidence.

As a second example we discuss how enhanced backscattering manifests itself
when electrons are injected into a Josephson junction. The system considered is
shown schematically in Fig.\ 3. A disordered metal grain is contacted by four
ballistic point contacts (with $N_{i}$ modes transmitted through contact
$i=1,2,3,4$). The scattering matrix $S$ has submatrices $s_{ij}$, the matrix
element $s_{ij,nm}$ being the scattering amplitude from mode $m$ in contact $j$
to mode $n$ in contact $i$. The grain forms a Josephson junction in a
superconducting ring. Coupling to the two superconducting banks is via point
contacts 3 and 4 (phase difference $\phi$, same electrostatic potential).
Contacts 1 and 2 are connected to normal metals (potential difference $V$). A
current $I$ is passed between contacts 1 and 2 and one measures the conductance
$G=I/V$ as a function of $\phi$. Spivak and Khmel'nitski\u{\i} computed
$\langle G(\phi)\rangle$ at temperatures higher than the Thouless energy
\cite{Spi82}. They found a periodic modulation of the weak-localization
correction, with amplitude of order $G_{0}$. Zaitsev and Kadigrobov et al.\
have discovered that at lower temperatures the amplitude increases to become
much greater than $G_{0}$ \cite{Zai94,Kad95}. Here we identify enhanced
backscattering as the origin of this increase.

The reflection matrices $r^{\rm ee}$ and $r^{\rm he}$ (with elements
$r_{ij,nm}$) contain the combined effect of scattering in the normal grain
(described by the matrix $S$) and Andreev reflection at the two contacts with
the superconductor. By summing a series of multiple Andreev reflections we
obtain expressions analogous to Eq.\ (\ref{rsrelation}),
\begin{mathletters}
\label{rsrelation2}
\begin{eqnarray}
&&r^{\rm ee}=a-b\,\Omega c^{\ast}\Omega^{\ast}(1+c\,\Omega
c^{\ast}\Omega^{\ast})^{-1}d,\label{ree2}\\
&&r^{\rm he}=-{\rm i}b^{\ast}\Omega^{\ast}(1+c\,\Omega
c^{\ast}\Omega^{\ast})^{-1}d,\label{rhe2}
\end{eqnarray}
\end{mathletters}%
where we have abbreviated
\begin{eqnarray*}
&&a=
{\renewcommand{\arraystretch}{0.6}
\left(\begin{array}{cc}
s_{11}&s_{12}\\s_{21}&s_{22}
\end{array}\right)},\;\;
b=
{\renewcommand{\arraystretch}{0.6}
\left(\begin{array}{cc}
s_{13}&s_{14}\\s_{23}&s_{24}
\end{array}\right)},\;\;
c=
{\renewcommand{\arraystretch}{0.6}
\left(\begin{array}{cc}
s_{33}&s_{34}\\s_{43}&s_{44}
\end{array}\right)},\nonumber\\
&&d=
{\renewcommand{\arraystretch}{0.6}
\left(\begin{array}{cc}
s_{31}&s_{32}\\s_{41}&s_{42}
\end{array}\right)},\;\;
\Omega=
{\renewcommand{\arraystretch}{0.6}
\left(\begin{array}{ll}
{\rm e}^{{\rm i}\phi/2}&0\\0&{\rm e}^{-{\rm i}\phi/2}
\end{array}\right)}.
\end{eqnarray*}
The four-terminal generalization of Eq.\ (\ref{GNS}) is \cite{Lam91}
\begin{mathletters}
\label{GRrelation}
\begin{eqnarray}
&&G/G_{0}=R_{21}^{\rm ee}+R_{21}^{\rm he}+\frac{2(R_{11}^{\rm he}R_{22}^{\rm
he}-R_{12}^{\rm he}R_{21}^{\rm he})}{R_{11}^{\rm he}+R_{22}^{\rm
he}+R_{12}^{\rm he}+R_{21}^{\rm he}},\label{GRrelationa}\\
&&R_{ij}^{\rm ee}={\textstyle\sum_{n,m}}|r_{ij,nm}^{\rm ee}|^{2},\;\;
R_{ij}^{\rm he}={\textstyle\sum_{n,m}}|r_{ij,nm}^{\rm
he}|^{2}.\label{GRrelationb}
\end{eqnarray}
\end{mathletters}%

Following Ref.\ \cite{Bar94}, we evaluate $\langle G\rangle$ by averaging $S$
over the circular ensemble. At $B=0$ this means that $S=UU^{\rm T}$ with $U$
uniformly distributed in the group ${\cal U}(M)$ of $M\times M$ unitary
matrices ($M=\sum_{i=1}^{4}N_{i}$). This is the circular orthogonal ensemble
(COE). If time-reversal symmetry is broken, then $S$ itself is uniformly
distributed in ${\cal U}(M)$. This is the circular unitary ensemble (CUE). In
the CUE we can do the average analytically for any $N_{i}$ and $\phi$. The
result is
\begin{equation}
\langle G\rangle_{\rm CUE}=G_{0}N_{1}N_{2}/(N_{1}+N_{2}),\label{GCUE}
\end{equation}
independent of $\phi$. In the COE we can do the average analytically for
$N_{i}\gg 1$ and $\phi=0$, and numerically \cite{Note2} for any $N_{i}$ and
$\phi$. We find that the difference $\Delta G(\phi)=\langle G(\phi)\rangle_{\rm
COE}-\langle G\rangle_{\rm CUE}$ is positive for $\phi=0$,
\begin{equation}
\frac{\Delta G(0)}{\langle G\rangle_{\rm CUE}}=
\rho+\case{1}{2}(1+\rho)^{2}-(1+\rho)\sqrt{\rho+\case{1}{4}(1+\rho)^{2}},
\label{DeltaG}
\end{equation}
with $\rho\equiv (N_{3}+N_{4})/(N_{1}+N_{2})$. The excess conductance
(\ref{DeltaG}) is a factor $G/G_{0}$ greater than the negative
weak-localization correction, which is observable in Fig.\ 3 at $\phi=\pi$. For
$N_{i}\gtrsim 10$ the finite-$N$ curves (solid) are close to the large-$N$
limit \cite{Note3} (dotted) which we have obtained using the Green's function
formulation of Refs.\ \cite{Naz94,Zai94}.

The excess conductance is a direct consequence of enhanced backscattering. This
is easiest to see for the symmetric case $N_{1}=N_{2}\equiv N$, when $\langle
R_{12}^{\rm he}\rangle=\langle R_{21}^{\rm he}\rangle$, $\langle R_{11}^{\rm
he}\rangle=\langle R_{22}^{\rm he}\rangle$. Current conservation requires
$R_{11}^{\rm he}+R_{21}^{\rm he}+R_{11}^{\rm ee}+R_{21}^{\rm ee}=N$. For $N\gg
1$ we may replace $\langle f(R_{ij})\rangle$ by $f(\langle R_{ij}\rangle)$. The
average of Eq.\ (\ref{GRrelation}) then becomes
\begin{equation}
\langle G/G_{0}\rangle=\case{1}{2}N-\case{1}{2}\langle R_{11}^{\rm
ee}-R_{21}^{\rm ee}\rangle+\case{1}{2}\langle R_{11}^{\rm he}-R_{21}^{\rm
he}\rangle.\label{Gsimplified}
\end{equation}
The first term $\case{1}{2}N$ is the classical series conductance. The second
term is the weak-localization correction due to enhanced backscattering for
normal reflection. Since $\langle R_{11}^{\rm ee}-R_{21}^{\rm ee}\rangle={\cal
O}(1)$ this negative correction to $\case{1}{2}N$ can be neglected if $N\gg 1$.
The third term gives the excess conductance due to enhanced backscattering for
Andreev reflection. Since $\langle R_{11}^{\rm he}-R_{21}^{\rm he}\rangle={\cal
O}(N)$ this positive contribution is a factor $G/G_{0}={\cal O}(N)$ greater
than the negative weak-localization correction.

In conclusion, we have predicted (and verified by numerical simulation) an
order $G/G_{0}$ enhancement of coherent backscattering by a disordered metal
connected to a superconductor. The enhancement can be observed as an excess
conductance which is a factor $G/G_{0}$ greater than the weak-localization
correction, provided ballistic point contacts are used to inject the current
into the junction. The junction should be sufficiently small that phase
coherence is maintained throughout. Several recent experiments \cite{exp} are
close to this size regime, and might well be equipped with ballistic point
contacts.

This work was supported by the Dutch Science Foundation NWO/FOM and by the
European Community.

\begin{figure}
\caption[]{
Numerical simulation of a $300\times 300$ tight-binding model for a disordered
normal metal ($L=9.5\,l$), in series with a superconductor (inset). The
histograms give the modal distribution for reflection of an electron at normal
incidence (mode number 1). The top two panels give the distribution of
reflected holes (for $B=0$ and $B=10\,h/eL^{2}$), the bottom panel of reflected
electrons (for $B=0$). The arrow indicates the ensemble-averaged height of the
backscattering peak for Andreev reflection, predicted from Eq.\
(\protect\ref{Andreev}).
}
\end{figure}

\begin{figure}
\caption[]{
Excess conductance $\Delta G=\langle G(B=0)\rangle -\mbox{}$ $\langle
G(B\gtrsim B_{\rm c})\rangle$ of a ballistic point contact in series with a
disordered NS junction (inset), computed from Eqs.\ (\protect\ref{GBzero}) and
(\protect\ref{GBc}). At $B=0$ the contact conductance is twice the Sharvin
conductance $N_{0}G_{0}$, provided $N_{0}L/Nl\ll 1$.
}
\end{figure}

\begin{figure}
\caption[]{Solid curves: excess conductance $\Delta G=\langle
G(\phi)\rangle_{\rm COE}$ $\mbox{}-\langle G\rangle_{\rm CUE}$ of a
four-terminal Josephson junction (inset), computed \protect\cite{Note2} from
Eqs.\ (\protect\ref{rsrelation2}) and (\protect\ref{GRrelation}) for
$N_{1}=N_{2}\equiv N$, $N_{3}=N_{4}\equiv\rho N$, with $N=10$. The dotted
curves are the large-$N$ limit \protect\cite{Note3}. The excess conductance at
$\phi=0$ is a factor $G/G_{0}={\cal O}(N)$ larger than the negative
weak-localization correction at $\phi=\pi$.
}
\end{figure}

\end{document}